\documentclass[aps,prd,amsmath,superscriptaddress,twocolumn,floatfix,nofootinbib,preprintnumbers]{revtex4-2}
\usepackage{amssymb}
\usepackage{txfonts}
\usepackage{epsfig}
\usepackage{bm}
\usepackage{color}
\usepackage{graphicx,graphics}
\usepackage{multirow}
\usepackage{float}
\usepackage{ulem}
\usepackage{setspace}
\usepackage{slashed}
\usepackage{booktabs}
\usepackage{hyperref}
\hypersetup{colorlinks, linkcolor = [rgb]{0, 0, 0.5}, citecolor = [rgb]{0,0.0,0.5}, urlcolor = [rgb]{0,0.0,0.5}}
\bibpunct{[}{]}{,}{n}{}{}
\allowdisplaybreaks[4]

\begin{document}

\title{Higher-twist effects in inclusive electron-positron annihilation}

\author{Jing Zhao}
\affiliation{Key Laboratory of Particle Physics and Particle Irradiation (MOE), Institute of Frontier and Interdisciplinary Science, Shandong University, \\Qingdao, Shandong 266237, China}

\author{Yongjie Deng}
\affiliation{Key Laboratory of Particle Physics and Particle Irradiation (MOE), Institute of Frontier and Interdisciplinary Science, Shandong University, \\Qingdao, Shandong 266237, China}

\author{Tianbo Liu}\thanks{Contact author:liutb@sdu.edu.cn}
\affiliation{Key Laboratory of Particle Physics and Particle Irradiation (MOE), Institute of Frontier and Interdisciplinary Science, Shandong University, \\Qingdao, Shandong 266237, China}
\affiliation{Southern Center for Nuclear-Science Theory (SCNT), Institute of Modern Physics, Chinese Academy of Sciences, Huizhou 516000, China}

\author{Weihua Yang}\thanks{Contact author:yangwh@ytu.edu.cn}
\affiliation{College of Nuclear Equipment and Nuclear Engineering, Yantai University,\\ Yantai, Shandong 264005, China}

\begin{abstract}

We establish a comprehensive theoretical framework for the electron-positron single-inclusive annihilation (SIA) process that incorporates higher-twist contributions up to twist-4 and investigate the possible relevance of these effects in low-energy reactions.
Through a systematic collinear expansion of the hadronic tensor, we derive the complete set of structure functions in terms of collinear fragmentation functions (FFs), explicitly including multiparton correlators up to the four-parton level.
To evaluate the possible impact of these power corrections, we estimate the twist-4 unpolarized FF $D_3$ using a spectator model and compute the normalized differential cross section for $\pi^0$ production. Our illustrative numerical estimation indicates that the interplay between kinematic hadron mass corrections and the higher-twist $D_3$ contribution can partially improve the description of recent BESIII data in the relatively low-$z$ region. The $Q$-dependence of this contribution is consistent with the expected power suppression of twist-4 terms. These findings suggest that higher-twist contributions are relevant for studies of hadronization dynamics at low energies and motivate more complete future analyses including all twist-4 FFs.

\end{abstract}

\maketitle

\section{Introduction}
\label{sec:intro}

Quantum chromodynamics (QCD) is the fundamental non-Abelian gauge theory of strong interactions. It is characterized by two essential features, i.e., asymptotic freedom and color confinement. Quarks and gluons participating high-energy reactions cannot be directly observed as asymptotic states, and only color-singlet hadrons can be detected. 
Understanding how color neutral hadrons are produced from colored quarks is one of the key issues in modern particle and nuclear physics. The process in which quarks and/or gluons combine to form color-singlet hadrons is known as hadronization. This process is inherently nonperturbative and cannot be calculated from first principles. Various phenomenological models were proposed to unravel the underlying mechanism of the hadronization process, such as the Feynman-Field model \cite{Field:1977fa}, the LUND model \cite{Andersson:1983ia}, the Webber model \cite{Marchesini:1983bm,Webber:1983if}, the recombination model \cite{Anisovich:1972pq,Bjorken:1973mh,Das:1977cp,Xie:1988wi} among others.
While these models provide intuitive physical pictures of the hadronization, they usually rely on process-dependent parameters, reducing the theoretical predictive power. The QCD factorization~\cite{Collins:1989gx} provides a theoretical framework, in which the hadronization is characterized by a set of universal and process-independent fragmentation functions (FFs), with the approximation control by the expansion in powers of the hard scale. At the leading power, one may attribute probability interpretation to the FFs, often referred to as leading-twist FFs.
For example, the $D_{f\rightarrow h}(z)$ represents the
probability density of a parton with flavor $f$ fragmenting into a hadron $h$ carrying the momentum fraction $z$. Hence, the nonperturbative transition from partons into hadrons is encoded in the FFs.

Fragmentation functions can be defined via the quark-quark and quark-gluon-quark correlation functions. In addition to leading-twist FFs, the decomposition of correlators can also yield higher-twist ones. While the contributions from higher-twist FFs are generally suppressed at high energy scales, they essentially encode the quantum interference and may have significant effects at lower-energy experiments, like JLab and BESIII, via power corrections~\cite{Liu:2019srj,Berger:1980qg}. In recent years, higher-twist effects have attracted considerable attention from theoretical perspectives. The complete sets of the higher-twist FFs have been derived via the parametrization of the correlation functions~\cite{Wei:2013csa,Yang:2017sxz,Bacchetta:2006tn,Wei:2014pma,Wei:2016far}.
The systematic framework for higher-twist calculations can be established through the collinear expansion method~\cite{Ellis:1982wd,Ellis:1982cd,Qiu:1990xxa,Qiu:1990xy}, which plays a central role in deriving the hadronic tensor in terms of gauge-invariant FFs in the annihilation process. The hard parts after the collinear expansion are perturbatively calculable and independent of the parton momenta except some delta-functions.
Correspondingly, the involved gauge-invariant FFs from the quark-quark or quark-gluon-quark correlators depend on one parton momentum. Hence the Lorentz decomposition of such quark-quark or quark-$j$-gluon-quark correlator is feasible and higher-twist calculations can be carried out~\cite{Wei:2013csa}.

Beyond the leading-power approximation, twist-4 contributions can be systematically organized in terms of gauge-invariant fragmentation correlators at the tree level, whereas their factorization and evolution beyond the tree level involve a more complex theoretical structure.
The relevant nonperturbative input involves multiparton correlation functions, which have been discussed in studies of power corrections and generalized factorization in DIS and related hard scattering processes~\cite{Ellis:1982wd,Ellis:1982cd,Qiu:1990xxa,Qiu:1990xy,Bartels:1999km,Jaffe:1983qp}. The renormalization of twist-4 operators is nontrivial~\cite{Kawamura:1996zr,Braun:2009mi}, and these operators mix with other multiparton and equation-of-motion operators under the scale evolution~\cite{Bukhvostov:1985rn,Ji:2014eta}. At subleading power, factorization studies in other processes provide concrete examples of the resulting evolution structure~\cite{Fleming:2013qu,Kang:2014tta}. In heavy quarkonium production, both single parton and double parton FFs enter the factorization formalism~\cite{KangQiuSterman:2012,Fleming:2012wy,Kang:2014tta,KangMaQiuSterman:2015,Ma:2014eja}, and their scale dependence is described by coupled evolution equations~\cite{Fleming:2013qu,Kang:2014tta,Lee:2020iim,Lee:2021oqr}. For SIDIS, factorization scale independence yields a double parton term suppressed by $1/\mu^2$ in the evolution of the single parton FF~\cite{Liu:2019srj}. Twist-4 factorization has also been established at NLO for this process~\cite{Kang:2014qba}. By contrast, existing parametrizations of twist-4 correlators for light hadron fragmentation in $e^+e^-$ annihilation have mainly been formulated at the tree level~\cite{Wei:2013csa,Chen:2016moq,Yang:2017sxz,Metz:2016swz}. The results available for DIS, SIDIS, and heavy quarkonium production do not yet provide a directly applicable renormalization and evolution framework for the specific light hadron twist-4 FFs entering the present SIA process. Accordingly, the formal derivation and numerical estimate in this work should be understood as a tree-level analysis at a fixed scale.

The hadron production in electron-positron single-inclusive annihilation (SIA) is the cleanest process to study the collinear FFs.
According to the factorization theorem~\cite{Collins:1989gx}, the cross section can be approximated as a convolution of the production of quarks and gluons at short distance and corresponding FFs to the desired hadron.
Since the FFs of light hadrons cannot be directly calculated in QCD from first principles, their determination mainly relies on the global analysis.
Over the past decades, the unpolarized leading-twist fragmentation functions 
are extensively studied. Most data are from measurements of pion and kaon productions in electron–positron annihilation experiments, such as those performed by  ALEPH~\cite{ALEPH:1994cbg,ALEPH:1992zhm}, Belle~\cite{Belle:2013lfg,Belle:2015hut}, \textit{BABAR}~\cite{BaBar:2013yrg} and BESIII~\cite{BESIII:2022zit,BESIII:2024hcs},
and have been implemented in global fits of leading-twist FFs by several groups, such as AKK~\cite{Albino:2005me,Albino:2005mv,Albino:2008fy}, DSS~\cite{deFlorian:2007aj,deFlorian:2007ekg}, MAPFF~\cite{AbdulKhalek:2022laj}, NNFF~\cite{Bertone:2017tyb}, and NPC~\cite{Gao:2024dbv,Gao:2024nkz}.
Recently, the measurements of inclusive $\pi^0$, $K_S^0$, and $\eta$ productions at a few GeV scales, $2\sim3.67\,{\rm GeV}$ were reported by BESIII collaboration~\cite{BESIII:2022zit,BESIII:2024hcs}.
A subsequent global analysis has been performed by NPC at next-to-next-to-leading order~\cite{Li:2024etc}. 
The fit results indicate the discrepancies between leading-twist predictions and BESIII experimental measurements. To describe these intermediate-energy scale data, they attribute the discrepancies to possible higher-twist contributions and add terms parametrized with higher-power scalings of $1/Q^2$.
For a full understanding of the fragmentation at these intermediate scales, it is required to establish the framework that incorporates higher-twist effects and enables a quantitative estimation of their contributions.

In this work, we aim to build up a systematic theoretical framework for SIA, including hadron mass corrections and dynamical higher-twist corrections. 
The paper is organized as follows. In Sec.~\ref{sec:formalism}, we derive the general form of the cross section for SIA process.
In Sec.~\ref{sec:partonmodel}, we calculate the cross section up to twist-4 in the parton model.
In Sec.~\ref{sec:model}, we perform the numerical estimates for normalized differential cross section including the hadron mass correction and the higher-twist effect.
Finally, we give a brief summary in Sec.~\ref{sec:summary}.

\section{General expression of the cross section} \label{sec:formalism}

\begin{figure}
  \centering
 \includegraphics[width=0.8\linewidth]{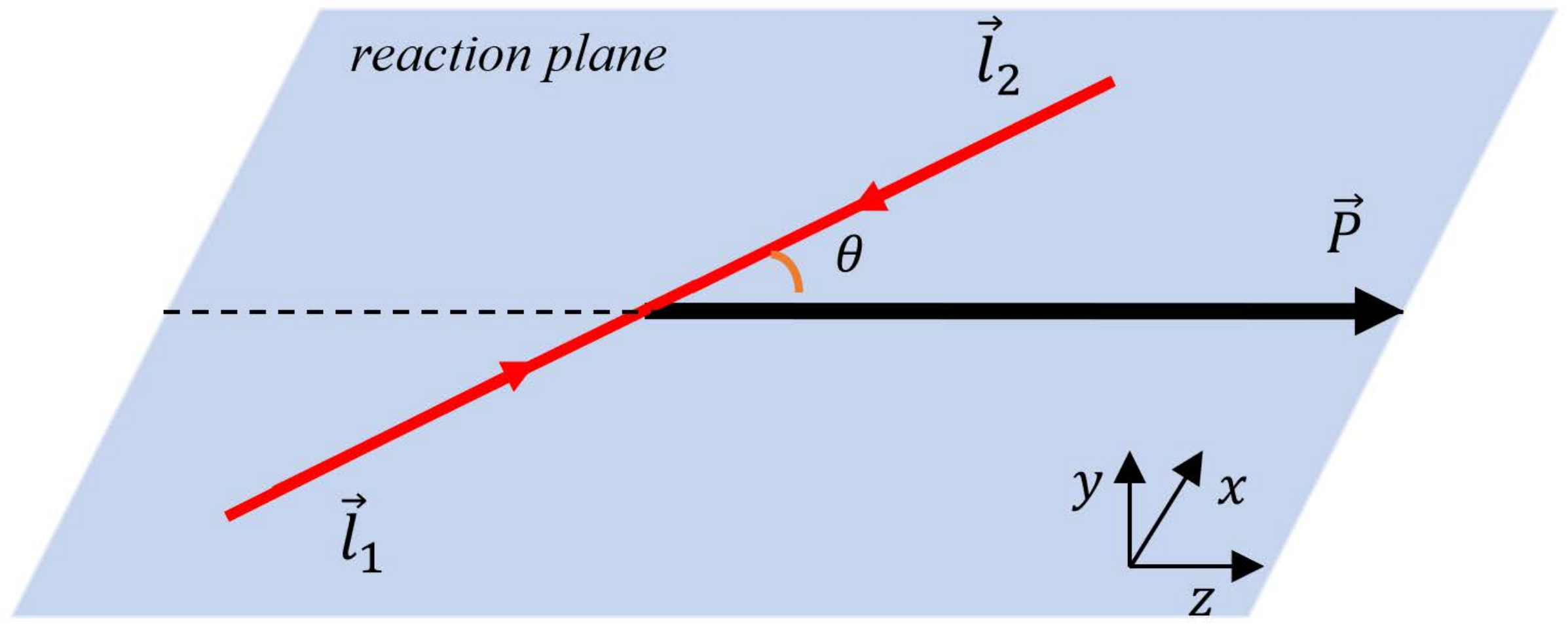}\\
  \caption{The center-of-mass frame of leptons of the inclusive electron positron annihilation process. The produced hadron travels along with $z$ direction.}\label{fig:frame}
\end{figure}

We consider the polarized electron-positron annihilation process, 
\begin{align}
    e^-(l_1) + e^+(l_2)\rightarrow h(P) +X(P_X), 
\end{align}
where four momenta of the relevant particles are labeled in parentheses. We denote the hadron by $h$, with mass $M$. 
The commonly used dimensionless variables are defined as
\begin{align}
  &y=\frac{P\cdot l_1}{P\cdot q}, &z=\frac{2P\cdot q}{Q^2},\label{e.variables}
\end{align}
where $Q^2=q^2=(l_1+l_2)^2$. As illustrated in Fig.~\ref{fig:frame}, we choose the center-of-mass frame of the leptons, where the detected hadron travels along the $z$ direction, and $\theta$ is the polar angle of the electron momentum $l_1$. The production plane, i.e., $x$-$z$ plane, is determined by the momenta of incoming electron and the produced hadron. 
In this frame, one can express the momenta of the involved particles as 
\begin{align}
    q^\mu&=Q\left(1,0,0,0\right),\\
    l_1^\mu&=\frac{Q}{2}\left(1,\sin\theta,0,\cos\theta \right),\\
    l_2^\mu&=\frac{Q}{2}\left(1,-\sin\theta,0,-\cos\theta \right),\\
    P^\mu&=\frac{zQ}{2}\left(1,0,0,\beta \right),    
\end{align}
where $\beta=\sqrt{1-\frac{4M^2}{z^2 Q^2}}$ is the mass correction factor.
The polar angle $\theta$ can be written in Lorentz invariant forms as
\begin{align}
    &\cos\theta=\frac{1-2y}{\beta}, \quad \quad
    &&\sin\theta=\sqrt{\frac{\beta^2-(2y-1)^2}{\beta^2}}.
\end{align}
If $h$ is a spin-$1/2$ hadron, one can also consider its polarization,
\begin{align}
    S^\mu=\left(S_L\frac{zQ\beta}{2M}, |S_T|\cos\phi_S, |S_T|\sin\phi_S, S_L\frac{zQ}{2M} \right),
\end{align}
where $S_L$ indicates the longitudinal polarization and $S_T$ denotes the transverse polarization of the hadron. For vector mesons and even higher-spin hadrons, one needs polarization tensors to fully characterize its spin. In this work, we only keep the polarization vector for simplicity.

With one-photon-exchange approximation, the differential cross section can be expressed as
\begin{align}
    \frac{{\rm d}\sigma}{{\rm d}z \,{\rm d}\cos\theta}=\frac{\pi\beta z \alpha^2}{2Q^4}L_{\mu\nu}W^{\mu\nu},
\end{align}
where $\alpha$ is the electromagnetic fine structure constant.
The leptonic tensor is given by
\begin{align}
L_{\mu \nu}(l_1, l_2)&=2\left[l_{1\mu} l_{2\nu} + l_{1\nu} l_{2\mu} -  g_{\mu \nu}(l_1 \cdot l_2)\right]+2i\lambda_e \varepsilon_{\mu\nu l_1 l_2},
\label{e.leptontensor}
\end{align}
where $\lambda_e$ denotes the helicity of the lepton. The hadronic tensor is given by
\begin{align}
W^{\mu \nu}\left(q ; P,S\right)=&\sum_X \left\langle 0\left|J^{\mu}(0)\right| P_{X} ; P , S\right\rangle\left\langle P_{X} ; P,  S\left|J^{\nu}(0)\right| 0\right\rangle \nonumber\\
&\times (2 \pi)^{3} \delta^{(4)}\left(q-P_{X}-P \right),\label{e.hadrontensor}
\end{align}
where $J^\mu$ is the electromagnetic current operator. 

Although one cannot directly calculate the hadronic tensor from first principles, its structure is constrained by required symmetries, the Hermiticity, gauge invariance, and parity invariance.
By imposing these constraints, one can construct five independent basis Lorentz tensors to decompose the hadronic tensor,
\begin{align}
h^{S\,\mu\nu}_{U} &= \left\{ g^{\mu\nu} - \frac{q^\mu q^\nu}{q^2},\; P_q^\mu P_q^\nu \right\}, \label{e.hus}\\
h_V^{S\,\mu\nu} &=\left\{ P_q^{\{\mu}\varepsilon^{\nu\}PqS} \right\}, \\
h_V^{A\,\mu\nu}&=\left\{(S\cdot q)\varepsilon^{\mu\nu qP }, P_q^{[\mu}\varepsilon^{\nu]PqS} \right\}, 
\end{align}
where $P_q^\mu\equiv P^\mu-q^\mu (P\cdot q)/q^2$, satisfying $P_q\cdot q=0$.
The subscripts $U$ and $V$ stand for the unpolarized and vector polarized hadron states, respectively.
The tensors with superscripts $S$ and $A$ are the symmetric and antisymmetric terms, respectively.
Given the basis Lorentz tensors above, the hadronic tensor is a linear combination of them,
\begin{align}
W^{\mu\nu} &= \sum_{i}^2 V^S_{U i} h^{S\,\mu\nu}_{U i} +\sum_{i}^1 V^S_{V i} h^{S\,\mu\nu}_{V i}+\sum_{i}^2 V_{Vi}^A h_{Vi}^{A\mu\nu} ,\label{e.Wmunu} 
\end{align}
where the coefficients $V$'s are scalar functions of $q^2$ and $P\cdot q$.

Contracting the leptonic tensor~\eqref{e.leptontensor} with the hadronic tensor~\eqref{e.Wmunu}, one can obtain the general expression of the differential cross section.
According to the angular distributions and the hadron spin states,
the differential cross section can be expressed as
\begin{align}
    \frac{{\rm d}\sigma}{{\rm d}z\,{\rm d}\cos\theta}=&\frac{\pi\beta z \alpha^2}{2Q^2} \bigg[
     (1+\cos^2\theta)F_{U,U}^T +\sin^2\theta F_{U,U}^L\nonumber\\ 
     &+|S_T|\sin2\theta \sin\phi_S F_{T,U}^{\sin\phi_S} +\lambda_e S_L \cos\theta F_{L,L}\nonumber\\
     &+\lambda_e |S_T| \sin\theta \cos\phi_S F_{T,L}^{\cos\phi_S}\bigg], \label{e.crosssf}
\end{align}
where the functions $F_{A,B}$'s are known as structure functions, depending on $z$ and $Q^2$.
The first subscript $A$ and the second subscript $B$ represent the spin states of the produced hadron and the lepton beam, respectively.
We use the superscript to indicate the azimuthal modulations.
For those independent azimuthal angles, we use superscripts $L$ and $T$ to distinguish the longitudinal and the transverse polarization of the virtual photon.

\section{Calculations in the parton model} \label{sec:partonmodel}

\subsection{Hadronic tensor and correlators} \label{sec:hadronictensor}

\begin{figure}
  \centering
 \includegraphics[width=1\linewidth]{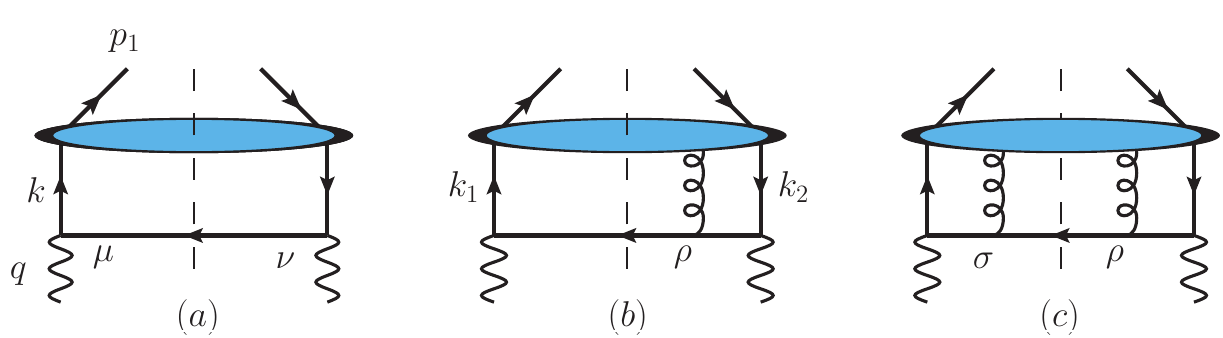}\\
  \caption{The first few diagrams as examples of the considered diagram series with exchange of $j$-gluon(s) and different cuts.
 We see (a) $j=0$, (b) $j=1$, and 
  (c) $j=2$, respectively.}\label{fig:feyn}
\end{figure}

We consider a series of diagrams, as illustrated in Fig.~\ref{fig:feyn}, which contain the exchange of $j$ gluon(s) in addition to the quark.
Here we truncate the series up to twist-4 and neglect the higher contributions for simplicity. After performing the collinear expansion, we can write the inclusive hadronic tensor as
\begin{align}
&W_{\mu\nu}(q,P)=\sum_{j,c}\tilde W^{(j,c)}_{\mu\nu}(q, P),\label{f:Wsum}
\end{align}
where $\tilde W^{(j,c)}_{\mu\nu}$ is a trace of the hard part and the quark-$j$-gluon-quark correlator. The superscript $c$ denotes different cuts. The explicit expressions can be simplified to~\cite{Wei:2014pma}
\begin{align}
   & \tilde W^{(0)}_{\mu\nu}
   = \frac{1}{2} \mathrm{Tr}\big[\hat h^{(0)}_{\mu\nu} \hat\Xi^{(0)}\big],\label{f:tW0}\\
   & \tilde W^{(1,L)}_{\mu\nu}
   =-\frac{1}{4(P\cdot q)} \mathrm{Tr}\big[\hat h^{(1)\rho}_{\mu\nu}\hat\Xi^{(1)}_{\rho} \big],\label{f:tW1L}\\
   & \tilde W^{(2,M)}_{\mu\nu}
   =\frac{1}{4(P\cdot q)^2} \mathrm{Tr}\big[\hat h^{(2)\rho\sigma}_{\mu\nu} \hat\Xi^{(2,M)}_{\rho\sigma} \big],\label{f:tW2M}\\
   & \tilde W^{(2,L)}_{\mu\nu}
   =\frac{1}{4(P\cdot q)^2} \mathrm{Tr}\big[ \hat N_{\mu\nu}^{(2)\rho\sigma} \hat\Xi^{(2)}_{\rho\sigma} +\hat h^{(1)\rho}_{\mu\nu}\hat\Xi^{(2\prime)}_{\rho} \big],\label{f:tW2L}
\end{align}
where we have omitted the dependence $(q,k_\perp^\prime)$ for simplicity. The hard parts are given by
\begin{align}
&\hat h^{(0)}_{\mu\nu} =\frac{1}{P^+} \gamma_\mu \slashed n \gamma_\nu , \label{f:h0}\\
&\hat h_{\mu\nu}^{(1)\rho} = \gamma_\mu \slashed n \gamma^\rho \slashed{\bar n}\gamma_\nu,  \label{f:h1}\\
&\hat N^{(2)\rho\sigma}_{\mu\nu} = q^- \gamma_\mu \gamma^\rho \slashed n \gamma^\sigma \gamma_\nu, \label{f:N2}\\
&\hat h_{\mu\nu}^{(2)\rho\sigma}=\frac{P^+}{2} \gamma_\mu \slashed{\bar n} \gamma^\rho \slashed n \gamma^\sigma \slashed{\bar n} \gamma_\nu.  \label{f:h2}
\end{align}
The quark-$j$-gluon-quark correlators are given by
\begin{align}
\hat \Xi^{(0)} =& \sum_X \int \frac{P^+d \xi^-}{2\pi} e^{-iP^+\xi^-/z}
  \langle 0 | \mathcal{L}^\dagger(0,\infty)\psi(0) |hX\rangle \nonumber\\
 &\times \langle hX| \bar\psi(\xi^-) \mathcal{L}(\xi^-,\infty) |0\rangle, \label{f:Xi0}\\
  \hat \Xi^{(1)}_{\rho}  = & \sum_X  \int\frac{P^+d\xi^-}{2\pi} e^{-iP^+\xi^-/z} \langle 0 | \mathcal{L}^\dagger (0,\infty) D_{\rho}(0) \psi(0) |hX\rangle \nonumber\\
&\times\langle hX| \bar\psi(\xi^-) \mathcal{L}(\xi^-,\infty) |0\rangle, \label{f:t3Xi1} \\
\hat \Xi^{(2,M)}_{\rho\sigma} = & \sum_X  \int\frac{P^+d\xi^-}{2\pi} e^{-iP^+\xi^-/z}  \langle 0 | \mathcal{L}^\dagger (0,\infty) D_{\rho}(0) \psi(0) |hX\rangle
\nonumber\\
& \times\langle hX| \bar\psi(\xi^-) D_\sigma (\xi^-)  \mathcal{L}(\xi^-,\infty) |0\rangle, \label{f:Xi2M}\\
\hat \Xi^{(2')}_{\rho} = & \sum_X \int\frac{P^+ d\xi^-}{2\pi} e^{-iP^+\xi^-/z} P^\sigma \langle 0 | \mathcal{L}^\dagger (0,\infty) D_{\rho}(0)  D_\sigma (0) \nonumber\\
&\times \psi(0) |hX\rangle\langle hX|\bar\psi(\xi^-) \mathcal{L}(\xi^-,\infty) |0\rangle, \label{f:Xi2B}\\
 \hat \Xi^{(2)}_{\rho\sigma}= & \sum_X  \int\frac{P^+d\xi^- }{2\pi}  e^{-iP^+\xi^-/z} \int iP^+ d\eta^- e^{-iP^+\eta^-/z} \nonumber\\
 &\langle 0 |\mathcal{L}^\dagger (\eta^-,\infty) D_{\rho}(\eta^-) D_{\sigma} (\eta^-)
 \nonumber \\
 &\times\mathcal{L}^\dagger (0,\eta^-)  \psi(0) |hX\rangle \langle hX| \bar\psi(\xi^-) \mathcal{L}(\xi^-,\infty) |0\rangle, \label{f:Xi2}
\end{align}
where $D_\rho=i\partial_\rho-gA_\rho$ denotes the transverse covariant derivative. $\mathcal{L}(0,y)$ is the gauge link, which is obtained by summing all $j$-gluon-exchange diagrams. 
The argument $\xi$ in the quark field operator $\psi$ and the gauge link represents $(0,\xi^-, \vec \xi_T)$.
We note that the leading-power contribution of $\tilde W^{(j)}_{\mu\nu}$ is of twist-$(j+2)$. Therefore, the second term in Eq.~(\ref{f:tW2L}) has no contribution up to twist-4 because of the factor $P^\sigma$ in the definition of $\hat\Xi^{(2')}_\rho$ given in Eq.~(\ref{f:Xi2B}).

In the SIA process, there is no helicity flip, which implies that only the chiral-even quantities contribute.
Therefore, we only need to consider the $\gamma^\alpha$- and the $\gamma^5\gamma^\alpha$-terms in the Dirac decomposition of the correlators, 
\begin{align}
    \hat \Xi^{(0)}=\frac{1}{2}\gamma^\alpha \Xi_\alpha^{(0)}+\frac{1}{2}\gamma^5\gamma^\alpha \tilde\Xi_\alpha^{(0)}. 
\end{align} 
Here $\Xi_\alpha^{(0)}$ and $\tilde \Xi_\alpha^{(0)}$ are coefficient functions which can be obtained through the following projections:
\begin{align}
 & \Xi_\alpha^{(0)} =\frac{1}{2}\mathrm{Tr}[\gamma^\alpha \hat\Xi^{(0)}], \label{f:projection1}\\
 & \tilde\Xi_\alpha^{(0)} =\frac{1}{2}\mathrm{Tr}[\gamma^\alpha\gamma^5 \hat\Xi^{(0)}].\label{f:projection2}
\end{align}
These coefficient functions can be further decomposed into fragmentation functions (FFs), as presented in Ref.~\cite{Goeke:2005hb}.
For the quark-quark correlator $\Xi^{(0)}$, we have
\begin{align}
  z\Xi^{(0)}_\alpha &=P^+\bar{n}_\alpha D_1 - M\tilde S_{T\alpha}D_T +\frac{M^2}{P^+}n_\alpha D_3,\label{f:xi0even}\\
  z\tilde\Xi^{(0)}_\alpha &=-P^+\bar{n}_\alpha S_L G_{1L} + M S_{T\alpha}G_T  -\frac{M^2}{P^+}n_\alpha S_L G_{3L}. \label{f:xi0odd}
\end{align}
Here, $\tilde S_{T\alpha}=\varepsilon_{\perp\alpha}^{S}=\varepsilon_{\perp\alpha}^{\beta}S_\beta$ with $\varepsilon_\perp^{\mu\nu}=\varepsilon^{\alpha\beta\mu\nu}\bar n_\alpha n_\beta$. For the quark-gluon-quark correlator $\Xi^{(1)}$, we have 
\begin{align}
  z\Xi^{(1)}_{\rho\alpha}&=-P^+\bar n_\alpha M\tilde S_{T\rho}D_{dT}+ M^2 g_{\perp\rho\alpha} D_{3d} + i S_L M^2\varepsilon_{\perp\rho\alpha} D_{3dL}, \label{f:xi1even3} \\
  z\tilde\Xi^{(1)}_{\rho\alpha}&=-iP^+\bar n_\alpha MS_{T\rho}G_{dT} +iM^2\varepsilon_{\perp\rho\alpha} G_{3d} + S_L M^2 g_{\perp\rho\alpha}G_{3dL}.\label{f:xi1odd3}
\end{align}
Here the subscript $d$ is used to denote FFs originated from quark-gluon-quark correlator. The transverse metric tensor is defined as $g_{\perp\mu\nu}= g_{\mu\nu}-\bar n_\mu n_\nu-\bar n_\nu n_\mu$.
For $\Xi^{(2)}$, we have 

\begin{align}
 z\Xi^{(2)}_{\rho\sigma\alpha} &= P^+\bar  n_\alpha 
   \left(M^2 g_{\perp\rho\sigma} D_{3dd} - i S_L M^2\varepsilon_{\perp\rho\sigma} D_{3dd L}\right), \label{f:Xi2decomp} \\
 z\tilde\Xi^{(2)}_{\rho\sigma\alpha} & = P^+\bar n_\alpha 
    \left(iM^2 \varepsilon_{\perp\rho\sigma} G_{3dd} -S_L M^2 g_{\perp\rho\sigma} G_{3dd L}\right). \label{f:tXi2decomp}
\end{align}
For readability, we summarize the FFs up to twist-4 obtained from the decompositions in Eqs.~\eqref{f:xi0even}--\eqref{f:tXi2decomp} in Table~\ref{tab:ff-summary}.

\begin{table*}
\caption{Summary of collinear FFs up to twist-4.}
\label{tab:ff-summary}
\begin{center}
\begin{tabular}{@{}lll@{}}
\hline\hline
\parbox[t]{0.34\textwidth}{Fragmentation functions} &
\parbox[t]{0.16\textwidth}{Twist} &
\parbox[t]{0.42\textwidth}{Originating correlator} \\
\hline
\parbox[t]{0.34\textwidth}{$D_1$, $G_{1L}$} & \parbox[t]{0.16\textwidth}{Twist-2} & \parbox[t]{0.42\textwidth}{Quark-quark correlator} \\
\parbox[t]{0.34\textwidth}{$D_T$, $G_T$} & \parbox[t]{0.16\textwidth}{Twist-3} & \parbox[t]{0.42\textwidth}{Quark-quark correlator} \\
\parbox[t]{0.34\textwidth}{$D_3$, $G_{3L}$} & \parbox[t]{0.16\textwidth}{Twist-4} & \parbox[t]{0.42\textwidth}{Quark-quark correlator} \\
\parbox[t]{0.34\textwidth}{$D_{dT}$, $G_{dT}$} & \parbox[t]{0.16\textwidth}{Twist-3} & \parbox[t]{0.42\textwidth}{Quark-gluon-quark correlator} \\
\parbox[t]{0.34\textwidth}{$D_{3d}$, $D_{3dL}$, $G_{3d}$, $G_{3dL}$} & \parbox[t]{0.16\textwidth}{Twist-4} & \parbox[t]{0.42\textwidth}{Quark-gluon-quark correlator} \\
\parbox[t]{0.34\textwidth}{$D_{3dd}$, $D_{3ddL}$, $G_{3dd}$, $G_{3ddL}$} & \parbox[t]{0.16\textwidth}{Twist-4} & \parbox[t]{0.42\textwidth}{Quark-gluon-gluon-quark correlator} \\
\parbox[t]{0.34\textwidth}{$D_{4q}$, $G_{4qL}$} & \parbox[t]{0.16\textwidth}{Twist-4} & \parbox[t]{0.42\textwidth}{Four-quark correlator} \\
\hline\hline
\end{tabular}
\end{center}
\end{table*}

Similarly, we use subscript $dd$ to denote FFs obtained from quark-gluon-gluon-quark correlator.
For correlator $\hat\Xi^{(2,M)}_{\rho\sigma}$, we require that its decomposition takes exactly the same form as that of $\hat\Xi^{(2)}_{\rho\sigma}$.
We distinguish them by adding an additional superscript $M$, and omit the explicit expressions.

According to Eqs. (\ref{f:xi0even})--(\ref{f:tXi2decomp}), we find that the decompositions of $\Xi$ and those of $\tilde \Xi$ have exact one-to-one correspondence: for each FF of type $D$, there exists a corresponding function of type $G$ and vice versa. Because of the Hermiticity of  $\hat{\Xi}^{(0)}$ and  $\hat{\Xi}^{(2,M)}_{\rho\sigma}$, the FFs defined via these correlators are real.  In contrast, for those FFs defined via $\hat{\Xi}_\rho^{(1)}$ and $\hat{\Xi}_{\rho\sigma}^{(2)}$, there is no such constraint so that they can be complex.

From the QCD equation of motion, $\gamma\cdot D\psi=0$, we can relate the quark-j-gluon-quark correlators to the quark-quark correlator. This implies that not all of the FFs introduced above are independent. For instance, the relations for the transverse components of the correlators  ${\Xi}_T^{(0)\rho}$ and $\tilde{\Xi}_T^{(0)\rho}$ can be written as
\begin{align}
  k^+\Xi_{T}^{(0)\rho}&=-g_\perp^{\rho\sigma} \mathrm{Re}\Xi^{(1)}_{\sigma +}-\varepsilon^{\rho\sigma}_{\perp}\mathrm{ Im} \tilde \Xi^{(1)}_{\sigma +},\label{eq:perpeom}\\
  k^+\tilde \Xi_{T}^{(0)\rho}&=-g_\perp^{\rho\sigma} \mathrm{Re}\tilde \Xi^{(1)}_{\sigma +}-\varepsilon^{\rho\sigma}_{\perp}\mathrm{ Im} \Xi^{(1)}_{\sigma +}. \label{eq:tperpeom}
\end{align}
Substituting the correlators given in Eqs. (\ref{f:xi0even})--(\ref{f:xi1odd3}) into Eqs. (\ref{eq:perpeom}) and (\ref{eq:tperpeom}), we obtain the following relations between twist-3 FFs,
\begin{align}
 &  D_T - iG_T =-z(D_{dT}-G_{dT}). \label{f:t3p}
\end{align}
In addition, the relations for the minus components of $\Xi^{(0)}_\alpha$ and $\tilde\Xi^{(0)}_\alpha$ are given by
\begin{align}
  2k^{+2}\Xi^{(0)}_- &=k^+\Big(g_\perp^{\rho\sigma} \Xi^{(1)}_{\rho\sigma } + i\varepsilon_\perp^{\rho\sigma}\tilde \Xi^{(1)}_{\rho\sigma }\Big) \nonumber\\
 &=- g_\perp^{\rho\sigma} \Xi^{(2,M)}_{\rho\sigma +}+i\varepsilon_\perp^{\rho\sigma}\tilde \Xi^{(2,M)}_{\rho\sigma +},\label{eq:minuseom}\\
  2k^{+2}\tilde\Xi^{(0)}_- &=k^+\Big(g_\perp^{\rho\sigma}\tilde \Xi^{(1)}_{\rho\sigma } + i\varepsilon_\perp^{\rho\sigma} \Xi^{(1)}_{\rho\sigma }\Big)\nonumber\\
 & =- g_\perp^{\rho\sigma}\tilde \Xi^{(2,M)}_{\rho\sigma +}+i\varepsilon_\perp^{\rho\sigma}\Xi^{(2,M)}_{\rho\sigma +}. \label{eq:tminuseom}
\end{align}
We can obtain the following equations
\begin{align}
 &  D_{3}  = z D_{-3d} = - z^2D_{+3dd}^{M}, \label{eq:t4eomD} \\
 &  G_{3L}  = z G_{-3dL} = -z^2G_{+3ddL}^{M}, \label{eq:t4eomG}
\end{align}
where 
\begin{align}
 & D_{-3d}\equiv D_{3d}-G_{3d}, \label{f:d-3}\\
 & G_{-3dL}\equiv G_{3dL}-D_{3dL}, \\
 & D_{+3dd}^M\equiv D_{3dd}^M+G_{3dd}^M, \\
 & G_{+3ddL}^{M}\equiv G_{3ddL}^{M}+D_{3ddL}^{M}. \label{f:g+3}
\end{align}
 These relations can be used to eliminate nonindependent higher-twist FFs.

\subsection{Leading-twist hadronic tensor}

We begin with the calculation of the leading-twist hadronic tensor.
The leading-twist contributions arise solely from the quark-quark correlator $\hat \Xi^{(0)}$. To calculate them, we use Eqs. (\ref{f:xi0even})--(\ref{f:xi0odd}) and the following relations
\begin{align}
  & \mathrm{Tr}\big[\hat h^{(0)}_{\mu\nu} \slashed {\bar n}\big]
 =-\frac{4}{P^+}g_{\perp\mu\nu},\label{f:traceh0} \\
  & \mathrm{Tr}\big[\hat h^{(0)}_{\mu\nu} \gamma^5 \slashed {\bar n}\big]
 =\frac{4}{P^+}i\varepsilon_{\perp\mu\nu}.\label{f:traceh05}
\end{align}
Substituting them into Eq. (\ref{f:tW0}) we obtain the leading-twist hadronic tensor,
\begin{align}
  z\tilde W_{t2\mu\nu} =& -g_{\perp\mu\nu}D_1 - iS_L\varepsilon_{\perp\mu\nu}G_{1L}. \label{f:Wt2munu}
\end{align}
The current conservation is satisfied, namely $q^\mu \tilde W_{t2\mu\nu}=q^\nu \tilde W_{t2\mu\nu}=0$.

\subsection{Twist-3 hadronic tensor}

The twist-3 contributions come from both the quark-quark correlator $\hat \Xi^{(0)}$ and the quark-gluon-quark correlator $\hat \Xi^{(1)}_\rho$. We first calculate the contributions from the quark-quark correlator $\hat \Xi^{(0)}$. In this case, we have
\begin{align}
  & \mathrm{Tr}\big[\hat h^{(0)}_{\mu\nu} \slashed {S}_T\big]
 =\frac{4}{P^+}S_{T\{\mu} n_{\nu\}},\label{f:tracet3h0} \\
  & \mathrm{Tr}\big[\hat h^{(0)}_{\mu\nu} \gamma^5 \slashed {S}_T\big]
 =\frac{4}{P^+}i\varepsilon^{S}_{\perp[\mu} n_{\nu]}. \label{f:traceht305}
\end{align}
Using Eqs. (\ref{f:xi0even})--(\ref{f:xi0odd}) and substituting them into Eq. (\ref{f:tW0}) we obtain
\begin{align}
  z\tilde W^{(0)}_{t3\mu\nu} =& -\frac{M}{P^+}\left(\varepsilon^S_{\perp\{\mu} n_{\nu\}}D_T - i \varepsilon^{S}_{\perp[\mu} n_{\nu]}G_T \right). \label{f:W0t3}
\end{align}
For the twist-3 contribution from the quark-gluon-quark correlator $\hat \Xi^{(1)}_\rho$, we have
\begin{align}
 & \mathrm{Tr}\big[\hat h^{(1)\rho}_{\mu\nu} \slashed{\bar n}\big]
 = -8 g_{\perp\mu}^{\rho} \bar n_\nu, \label{f:tracet3h1}\\
 & \mathrm{Tr}\big[\hat h^{(1)\rho}_{\mu\nu}\gamma^5 \slashed{\bar n}\big]
  =-8i \varepsilon_{\perp\mu}^\rho \bar{n}_\nu. \label{f:traceht3h51}
\end{align}
Using Eqs. (\ref{f:xi1even3})--(\ref{f:xi1odd3}) and substituting them into Eq. (\ref{f:tW1L}), we obtain
\begin{align}
  z\tilde W^{(1)L}_{t3\mu\nu} = -\frac{M}{q^-}\left(\varepsilon_{\perp\mu}^S\bar{n}_\nu D_{dT}-\varepsilon_{\perp\mu}^S\bar{n}_\nu G_{dT} \right).\label{f:W1t3Ld}
\end{align}

The complete twist-3 hadronic tensor is the sum of all the twist-3 contributions, i.e., $\tilde W_{t3\mu\nu}=\tilde W^{(0)}_{t3\mu\nu}+\tilde W^{(1)L}_{t3\mu\nu}+\left(\tilde W^{(1)L}_{t3\nu\mu}\right)^*$.
By employing Eqs.~\eqref{f:t3p},~\eqref{f:W0t3}, and~\eqref{f:W1t3Ld}, we eliminate the nonindependent FFs and obtain the complete hadronic tensor at twist-3,
\begin{align}
  z\tilde W_{t3\mu\nu} = -\frac{M}{P\cdot q}\left(\varepsilon^S_{\perp\{\mu} \bar{q}_{\nu\}}D_T + i \varepsilon^S_{\perp[\mu} \bar{q}_{\nu]}G_T \right), \label{f:wt3munu}
\end{align}
where $\bar q =q -2P/z$.
We can verify that $\tilde W_{t3\mu\nu}$ satisfies the current conservation $q^\mu \tilde W_{t3\mu\nu}=q^\nu \tilde W_{t3\mu\nu}=0$.

\subsection{Twist-4 hadronic tensor}
The correlators $\hat \Xi^{(0)}$, $\hat \Xi^{(1)}_\rho$, and $\hat \Xi^{(2)}_{\rho\sigma}$ all contribute to the twist-4 hadronic tensor. We first compute the contributions from quark-quark correlator $\hat \Xi^{(0)}$ by using
\begin{align}
  & \mathrm{Tr}\big[\hat h^{(0)}_{\mu\nu} \slashed n\big]
 =\frac{8}{P^+}n_\mu n_\nu,\label{f:traceh0t4} \\
  & \mathrm{Tr}\big[\hat h^{(0)}_{\mu\nu} \gamma^5 \slashed n\big]
 =0,\label{f:traceh05t4}
\end{align}
and Eqs.~\eqref{f:xi0even}--\eqref{f:xi0odd}. Substituting the twist-4 terms into Eq.~\eqref{f:tW0}, we obtain
\begin{align}
  z\tilde W^{(0)}_{t4\mu\nu} =& \frac{2M^2}{(P^+)^2} n_\mu n_\nu D_3. \label{f:W0t4munu}
\end{align}

To calculate the twist-4 contributions from quark-gluon-quark correlator $\hat \Xi^{(1)}_\rho$, we use
\begin{align}
   \mathrm{Tr}\big[\hat h^{(1)\rho}_{\mu\nu} \gamma^\alpha \big]&=4\big( 2n_\mu \bar n_\nu g_\perp^{\rho\alpha}+g_{\perp\mu\nu}g_\perp^{\rho\alpha} -g_{\perp\mu}^{\{\rho}g_{\perp\nu}^{\alpha\}}\big),\label{f:traceh1t4} \\
   \mathrm{Tr}\big[\hat h^{(1)\rho}_{\mu\nu} \gamma^5 \gamma^\alpha \big]&=4i\big(2n_\mu \bar n_\nu \varepsilon_\perp^{\rho\alpha}+g_{\perp\mu}^{~~\rho} \varepsilon_{\perp\nu}^{~~\alpha} +g_{\perp\nu}^{~~\alpha} \varepsilon_{\perp\mu}^{~~\rho}\big). \label{f:traceh15t4}
\end{align}
Using twist-4 FFs given in Eqs. (\ref{f:xi1even3})--(\ref{f:xi1odd3}) and Eq. (\ref{f:tW1L}), we have
\begin{align}
  z\tilde W^{(1)L}_{t4\mu\nu} =& -\frac{2M^2}{P\cdot q}n_\mu \bar n_\nu \left(D_{3d}-G_{3d}\right). \label{f:W1t4munu}
\end{align}

It is convenient to divide the contributions from quark-gluon-gluon-quark correlator $\hat \Xi^{(2)}_{\rho\sigma}$ into two parts, one is the middle-cut part and the other is the left- and right-cut part. We first consider the middle-cut case, for which we introduce the superscript $M$ to distinguish it from the others. Utilizing the relations
\begin{align}
  & \mathrm{Tr}\big[\hat h^{(2)\rho\sigma}_{\mu\nu} \slashed{\bar n} \big]P^+
 =-8 P_{\mu} P_{\nu}g_\perp^{\rho\sigma}, \label{f:traceh2t4} \\
  & \mathrm{Tr}\big[\hat h^{(2)\rho\sigma}_{\mu\nu} \gamma^5 \slashed{\bar n} \big]P^+=8i P_{\mu} P_{\nu}\varepsilon_\perp^{\rho\sigma},\label{f:traceh25t4}
\end{align}
we have
\begin{align}
  z\tilde W^{(2)M}_{t4\mu\nu} =& -\frac{2M^2}{(P\cdot q)^2} P_{\mu} P_{\nu} \left(D^M_{3dd}+G^M_{3dd}\right). \label{f:W2t4munuM}
\end{align}

To obtain the contributions from the left-cut and right-cut parts, we use
\begin{align}
  \mathrm{Tr}\big[\hat N^{(2)\rho\sigma}_{\mu\nu}\slashed {\bar n} \big]
  =&\frac{4(P\cdot q)}{P^+}\Big(g_\perp^{\rho\sigma}g_{\perp\mu\nu}+g_{\perp[\mu}^{~~\rho} g_{\perp\nu]}^{~~\sigma}\Big),  \label{f:traceh2Le}\\
  \mathrm{Tr}\big[\hat N^{(2)\rho\sigma}_{\mu\nu}\gamma^5\slashed {\bar n} \big]
  =&i\frac{4(P\cdot q)}{P^+}\Big(g_{\perp\mu}^{~~\rho}\varepsilon_{\perp\nu}^{~~\sigma} - g_{\perp\nu}^{~~\sigma}\varepsilon_{\perp\mu}^{~~\rho}\Big), \label{f:traceh2Lo}
\end{align}
and obtain
\begin{align}
  z\tilde W^{(2)L}_{t4\mu\nu} =& \frac{M^2}{P\cdot q}g_{\perp\mu\nu}\left(D_{3dd}-G_{3dd}\right)+ i\frac{M^2}{P\cdot q}\varepsilon_{\perp\mu\nu} S_L\left(D_{3ddL}-G_{3ddL}\right). \label{f:W2t4munuL}
\end{align}
Summing over all twist-4 contributions and utilizing Eqs. (\ref{eq:t4eomD})--(\ref{eq:t4eomG}) to eliminate the nonindependent FFs yields
\begin{align}
  z\tilde W_{t4\mu\nu}=& \frac{2M^2}{(P\cdot q)^2} \bar q_\mu \bar q_\nu D_3 
  + \frac{2M^2}{P\cdot q} g_{\perp\mu\nu} \left(D_{3dd}-G_{3dd}\right)\nonumber\\
  &+ i\frac{2M^2}{P\cdot q}\varepsilon_{\perp\mu\nu} S_L\left(D_{3ddL}-G_{3ddL}\right). \label{f:wt4munu}
\end{align}
It can be shown that $\tilde W_{t4\mu\nu}$ satisfies the current conservation, i.e. $q^\mu \tilde W_{t4\mu\nu}=q^\nu \tilde W_{t4\mu\nu}=0$.

\subsection{Four-quark correlator contributions}

\begin{figure}
  \centering
 \includegraphics[width=1\linewidth]{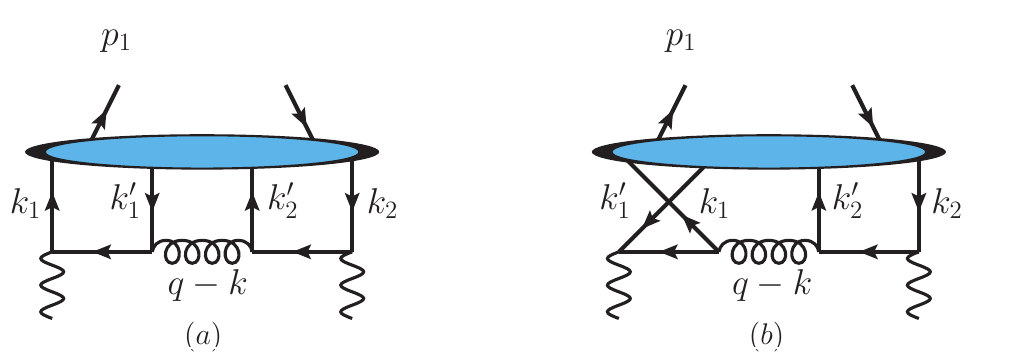}
  \caption{The first two diagrams correspond to the four-quark case without multiple gluon scattering.
  In (a), we have $k_1'=k_1-k$ and $k_2'=k_2-k$;
  in (b), we have the interchange of $k_1$ with $k_1'$. }\label{fig:feyn4q}
\end{figure}

In the previous calculations, we only consider the contributions from quark-j-gluon-quark correlators.
In fact, up to twist-4, there are also contributions from diagrams involving the four-quark correlator \cite{Qiu:1988dn}, which is defined by 
\begin{align}
  \hat \Xi^{(0)}_{(4q)}(k_1,k,k_2)=&\frac{g^2}{8}\sum_X\int\frac{d^4y}{(2\pi)^4}\frac{d^4y_1}{(2\pi)^4}\frac{d^4y_2}{(2\pi)^4} \nonumber\\
  &\times e^{-ik_1y+i(k_1-k)y_1-i(k_2-k)y_2} \nonumber\\
  &\times \langle 0|\bar\psi(y_2)\mathcal{L}^\dag(0,y_2)\psi(0)|hX \rangle\nonumber\\
  &\times \langle hX|\bar\psi(y)\mathcal{L}(y,y_1)\psi(y_1)|0 \rangle. \label{eq:ch2-4qcorr}
\end{align}
Considering diagrams in Fig.~\ref{fig:feyn4q}, we can write the hadronic tensors for both the quark jet or gluon jet \footnote{Quark jet or gluon jet represents the positions that cut lines go through in Fig. \ref{fig:feyn4q}. If the cut line is in the middle position, it is a gluon jet case.} cases in a unified form, 
\begin{align}
  \hat W_{4q\mu\nu}^{(g/q)}=\frac{1}{2P\cdot q} &\int dzdz_1dz_2  h^{g/q}_{4q}\Big(g_{\perp\mu\nu}\hat C_s+i \varepsilon_{\perp\mu\nu}\hat C_{ps}\Big). \label{eq:ch2-W4q-uni}
\end{align}
Here $\hat C_s$ and $\hat C_{ps}$ are the correlators and defined as
\begin{align}
  \hat C_j=&\int\frac{d^2k'_\perp}{(2\pi)^2}\int d^4k_1d^4kd^4k_2\delta\left(z-\frac{P^+}{k^+}\right)\nonumber\\
  &\times\delta\left(k_1^+ z_1-P^+\right)\delta\left(k_2^+ z_2-P^+\right)\nonumber\\
  &\times (2\pi)^2\delta^2(\vec k_\perp + \vec k_\perp')\hat \Xi^{(0)}_{(4q)j}(k_1,k,k_2;P,S), \label{eq:ch2-Cs-ps}
\end{align}
where subscript $j=s, ps$. The corresponding correlators $\hat \Xi^{(0)}_{(4q)s}$  and $\hat \Xi^{(0)}_{(4q)ps}$ are defined as 
\begin{align}
  \hat \Xi^{(0)}_{(4q)s}&=\frac{g^2}{8}\int\frac{d^4y}{(2\pi)^4}\frac{d^4y_1}{(2\pi)^4}\frac{d^4y_2}{(2\pi)^4} e^{-ik_1y+i(k_1-k)y_1-i(k_2-k)y_2} \nonumber\\
  &\sum_X \Big\{\langle 0|\bar\psi(y_2)\slashed n\psi(0)|hX \rangle \langle hX|\bar\psi(y)\slashed n\psi(y_1)|0 \rangle \nonumber\\
   &+\langle 0|\bar\psi(y_2)\gamma^5\slashed n\psi(0)|hX \rangle \langle hX|\bar\psi(y)\gamma^5\slashed n\psi(y_1)|0 \rangle  \Big\}, \label{eq:ch2-Xis}\\
  \hat \Xi^{(0)}_{(4q)ps}&=\frac{g^2}{8}\int\frac{d^4y}{(2\pi)^4}\frac{d^4y_1}{(2\pi)^4}\frac{d^4y_2}{(2\pi)^4} e^{-ik_1y+i(k_1-k)y_1-i(k_2-k)y_2} \nonumber\\
  &\sum_X \Big\{\langle 0|\bar\psi(y_2)\gamma^5\slashed n\psi(0)|hX \rangle \langle hX|\bar\psi(y)\slashed n\psi(y_1)|0 \rangle \nonumber\\
   &+\langle 0|\bar\psi(y_2)\slashed n\psi(0)|hX \rangle \langle hX|\bar\psi(y)\gamma^5\slashed n\psi(y_1)|0 \rangle  \Big\}. \label{eq:ch2-Xips}
\end{align}
For the sake of simplicity, we have neglected gauge links here. In Eq. (\ref{eq:ch2-W4q-uni}), the $h^{g/q}_{4q}$ denotes the sum of all hard scattering amplitude. For the gluon jet case, we have 
\begin{align}
  h_{4q}^{g}&=\frac{z z_B^3\delta(z-z_B)}{\big(z_1-z_B+i\epsilon\big)\big(z_2-z_B-i\epsilon\big)}  \nonumber\\
  & +\frac{z z_B^3\delta(z-z_B)}{\big(z_1+z_B-\frac{z_1z_B}{z}+i\epsilon\big)\big(z_2+z_B-\frac{z_2z_B}{z}-i\epsilon\big)} \nonumber\\
  & - \frac{z z_B^3\delta(z-z_B)}{\big(z_1-z_B+i\epsilon\big)\big(z_2+z_B-\frac{z_2z_B}{z}-i\epsilon\big)}\nonumber\\
  & - \frac{z z_B^3\delta(z-z_B)}{\big(z_1+z_B-\frac{z_1z_B}{z}+i\epsilon\big)\big(z_2-z_B-i\epsilon\big)}. 
\end{align}
Here $z$, $z_1$, $z_2$ denote the momentum fractions, while $z_B$ represents the variable as defined in Eq.~\eqref{e.variables}.
For the quark jet case, we have $h_{4q}^q=h_{4q}^{qL}+h_{4q}^{qR}$ with
\begin{align}
  h_{4q}^{qL}&=\frac{z z_B^3 \delta(z_1-z_B)}{\big(z-z_B-i\epsilon\big)\big(z_2-z_B-i\epsilon\big)}   \nonumber\\
  & + \frac{z z_B^3\delta(z_1+z_B-\frac{z_1z_B}{z})}{\big(z-z_B-i\epsilon\big)\big(z_2+z_B-\frac{z_2z_B}{z}-i\epsilon\big)}  \nonumber\\
  &- \frac{z z_B^3 \delta(z_1-z_B)}{\big(z-z_B-i\epsilon\big)\big(z_2+z_B-\frac{z_2z_B}{z}-i\epsilon\big)}  \nonumber\\
  & - \frac{z z_B^3 \delta(z_1+z_B-\frac{z_1z_B}{z})}{\big(z-z_B-i\epsilon\big)\big(z_2-z_B-i\epsilon\big)}.
\end{align}
Here $h_{4q}^{qR}$ can be obtained by $h_{4q}^{qR}(z_1,z,z_2)=h_{4q}^{qL*}(z_2,z,z_1)$. Summing over all contributions, we obtain $h_{4q}=h_{4q}^{qL}+h_{4q}^{qR}+h_{4q}^g$.

To derive an explicit expression, we decompose the correlators as 
\begin{align}
  z\int dzdz_1dz_2 & h_{4q}\hat C_{s}=M^2 D_{4q},\label{eq:ch2-cs4q}\\
  z\int dzdz_1dz_2 & h_{4q}\hat C_{ps}=S_L M^2 G_{4qL}, \label{eq:ch2-ps4q}
\end{align}
and obtain the four-quark correlator contribution
\begin{align}
  z W_{4q\mu\nu}=\frac{M^2}{(2P\cdot q)}\Big(g_{\perp\mu\nu}D_{4q}
  +iS_L \varepsilon_{\perp\mu\nu} G_{4qL} \Big).\label{f:w4qmunu}
\end{align}

Up to now, we have derived the complete hadronic tensor up to twist-4 level,
\begin{align}
     z W_{\mu\nu}=&-g_{\perp\mu\nu}D_1 - iS_L\varepsilon_{\perp\mu\nu}G_{1L}\nonumber\\
    &-\frac{M}{P\cdot q}\left(\varepsilon^S_{\perp\{\mu} \bar{q}_{\nu\}}D_T + i \varepsilon^S_{\perp[\mu} \bar{q}_{\nu]}G_T \right)\nonumber\\
    &+\frac{2M^2}{(P\cdot q)^2} \bar q_\mu \bar q_\nu D_3 
    + \frac{2M^2}{P\cdot q} g_{\perp\mu\nu} D_{-3dd}+ i\frac{2M^2}{P\cdot q}\varepsilon_{\perp\mu\nu} S_L D_{-3ddL}\nonumber\\
    &+\frac{M^2}{(2P\cdot q)}\Big(g_{\perp\mu\nu}D_{4q}
  +iS_L \varepsilon_{\perp\mu\nu} G_{4qL} \Big).\label{e.totalwmunu}
\end{align}
where the leading-twist part is given in Eq.~\eqref{f:Wt2munu}, the twist-3 contribution is presented in Eq.~\eqref{f:wt3munu}, the twist-4 terms are shown in Eqs.~\eqref{f:wt4munu} and~\eqref{f:w4qmunu}, $D_{-3dd}=D_{3dd}-G_{3dd}$. We note that all of these hadronic tensors satisfy the current conservation.

\subsection{Differential cross section up to twist-4}

The differential cross section is obtained through the contraction of the leptonic tensor and the hadronic tensor,
\begin{align}
    \frac{d\sigma}{dz d\cos\theta}=& \sum_a e_a^2 \frac{\pi\beta  \alpha^2 }{2Q^2} \Bigg\{
    (1+\cos^2\theta)\nonumber\\
    &\times\left(D_1 -\frac{4\kappa_M^2}{z}D_{-3dd}-\frac{\kappa_M^2}{z}D_{4q}\right)\nonumber\\
    &+\beta^2 \frac{8\kappa_M^2}{z^2}\sin^2\theta D_3 \nonumber\\
    &+2\lambda_e S_L \cos\theta\left(G_{1L}-\frac{\kappa_M^2}{z}G_{4qL}-\frac{4\kappa_M^2}{z}D_{-3ddL}\right)\nonumber \\
    &-|S_T|\beta \frac{\kappa
    _M}{z}\sin2\theta \sin\phi_S D_T\nonumber\\
    &-2\lambda_e|S_T|\beta\frac{\kappa_M}{z}\sin\theta \cos\phi_S G_T 
    \Bigg\}.\label{f:crossparton4}
\end{align}
The sum of ``$a$" runs over all active quark and antiquark flavors with $e_a$ being the number of electric charge.
A subscript $a$ should be explicitly added to the FFs, but here we drop it for simplicity.
The $\kappa_M\equiv M/Q$ is a typical suppression factor of higher-twist contributions. One can find that all twist-4 terms contain the factor $\kappa_M^2$.

Comparing this expression with that in Eq. (\ref{e.crosssf}), we can express the structure functions in terms of the FFs,
\begin{align}
    F_{U,U}^T&={\frac 1 z} \sum_a e_a^2 \left[D_{1} -\frac{4\kappa_M^2}{z}D_{-3dd}-\frac{\kappa_M^2}{z}D_{4q}\right],\\
    F_{U,U}^L&={\frac 1 z}\sum_a e_a^2\beta^2 \frac{8\kappa_M^2}{z^2} D_3,\\
    F_{T,U}^{\sin\phi_S}&=-{\frac 1 z} \sum_a e_a^2 \beta \frac{\kappa
    _M}{z}D_T,\\
    F_{L,L}&={\frac 2 z} \sum_a  e_a^2 \left(G_{1L}-\frac{\kappa_M^2}{z}G_{4qL}-\frac{4\kappa_M^2}{z}D_{-3ddL}\right),\\
    F_{T,L}^{\cos\phi_S}&=-{\frac 2 z} \sum_a e_a^2 \beta\frac{\kappa_M}{z} G_T,
\end{align}
It is interesting to note that $F_{U,U}^L$ is independent of $F_{U,U}^T$ and receives a twist-4 contribution only from $D_3$. Generally speaking, leading-twist FFs appearing in the unpolarized structure function acquire corrections from twist-4 FFs which come from the four-quark correlators and/or quark-gluon-gluon-quark correlators.

\section{Numerical estimates}\label{sec:model}

Since global analysis of higher-twist FFs is not yet available, we employ the spectator model in this section to provide an illustrative estimate of the $D_3$ contribution in the SIA process.
Based on the recent measurements of various final-state hadrons reported by the BESIII Collaboration~\cite{BESIII:2022zit,BESIII:2024hcs}, we focus primarily on the production of $\pi^0$ in this work. Moreover, the method is also applicable to other hadrons with proper parameters.

To describe the matrix elements in the correlator, we assume that the interaction is implemented through a pseudoscalar pion-quark coupling. The interaction Lagrangian can be written as~\cite{Bacchetta:2001di}
\begin{align}
    \mathcal{L}(x)=ig_q \bar{q}(x)\gamma_5 q(x)\pi(x).
\end{align}
Under this assumption, the fragmentation correlator~\eqref{f:Xi0} for the case $q\rightarrow \pi^0$ at tree level can be expressed as~\cite{Bacchetta:2007wc}
\begin{align}
\Xi^{(0)}(k, P) =& -\frac{2 g_{q}^2}{(2\pi)^4} \frac{(\slashed{k} + m)}{k^2 - m^2} \gamma_5 (\slashed{k} - \slashed{P} + m_s) \gamma_5 \nonumber \\ &\times \frac{(\slashed{k} + m)}{k^2 - m^2} \, 2\pi \, \delta\left((k - P)^2 - m_s^2\right),
\end{align}
where $m_s$ indicates the mass of the spectator quark.
Integrating over the minus component $k^-$, we obtain the transverse momentum dependent correlator,
\begin{align}
\Xi^{(0)}(z, \boldsymbol{k}_T) = \frac{2 g_{q}^2}{32\pi^3} 
\frac{(\slashed{k} + m)(\slashed{k} - \slashed{P} - m_s)(\slashed{k} + m)}{(1 - z) P^+ (k^2 - m^2)^2},
\end{align}
where $k^2$ is defined by
\begin{align}
    k^2=\frac{z\bm{k}_T^2}{1-z}+\frac{m_s^2}{1-z}+\frac{M^2}{z}.
\end{align}

As shown in Eqs.~\eqref{f:projection1} and~\eqref{f:projection2}, the TMD FFs can be projected out by the corresponding Dirac matrices
\begin{align}
    \Xi^{(0)[\Gamma]}(z, \boldsymbol{k}_T)=\frac{1}{2}\mathrm{Tr}[\Xi^{(0)}(z, \boldsymbol{k}_T)\Gamma].
\end{align}
This leads to the leading-twist and twist-4 FFs in the spectator model,
\begin{align}
    D_1(z,k_T^2) &=\frac{g_q^2}{8\pi^3}\frac{[z^2\bm{k}_T^2+(zm+m_s-m)^2]}{z^3(\bm{k}_T^2+L^2)^2}, \\
    D_3(z,k_T^2)
    &= \frac{g_q^2 }{16 \pi^3}\frac{\left(z^2 k_T^2 (m - m_s)^2 + N^4\right)}{ M^2 z^3 (k_T^2 + L^2)^2}. \label{f:D3}
\end{align}
Here $N^2$ and $L^2$ are defined as
\begin{align}
    N^2&=M^2(z-1)+zm_s(m-m_s), \\
    L^2&=\frac{1-z}{z^2}M^2+m^2+\frac{m_s^2-m^2}{z}.
\end{align}
After integrating the transverse momentum,
we can obtain the collinear unpolarized FF,
\begin{align}
    D_1(z)=\pi z^2 \int_0^\infty dk_T^2 D_1(z,k_T^2).
\end{align}
The twist-4 collinear unpolarized FF $D_3$ has a similar definition.
This integration over $\bm{k}_T$ is divergent at large $|\bm{k}_T|$.
In this work, we choose the Gaussian regularization as
\begin{align}
	g_q(k^2) \to  \dfrac{g_q}{z} e^{-\frac{k^2}{\Lambda^2}}\label{e.formfactor}
\end{align}
with $\Lambda^2=\lambda^2 z^{a}(1-z)^{b}$.

\begin{figure}
  \centering
 \includegraphics[width=0.8\linewidth]{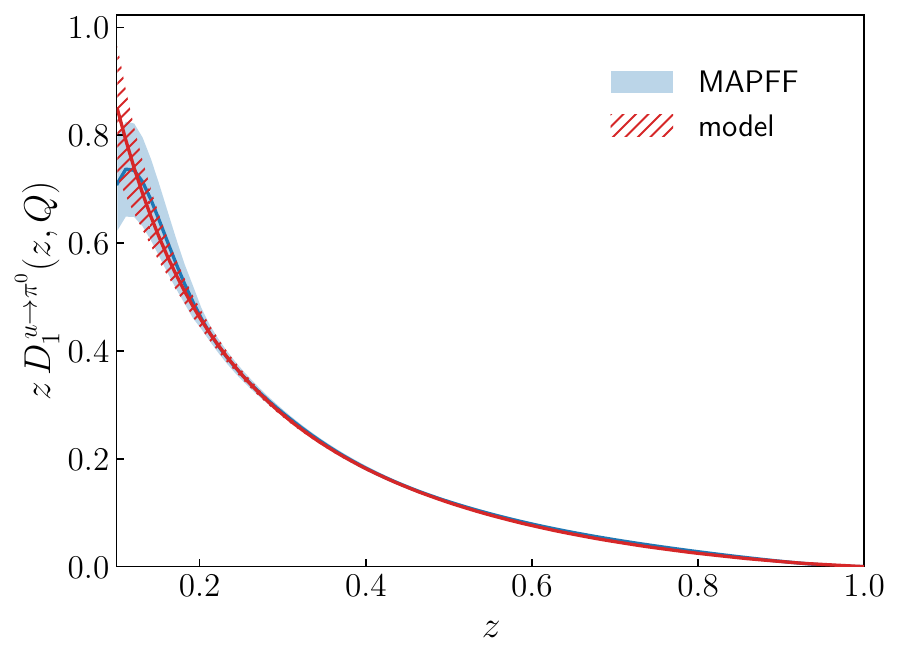}\\
  \caption{Comparison of the spectator model fit for $D_1$ with the MAPFF replica set at $Q=2.8$ GeV.}\label{fig:fitD1-MAPFF}
\end{figure}

Applying the isospin symmetry and the charge-conjugation, we obtain the following relations
\begin{align}    D_1^{u\to\pi^0}=D_1^{\bar{d}\to\pi^0}=D_1^{d\to\pi^0}=D_1^{\bar{u}\to\pi^0}.
\end{align}
We set the constituent quark mass $m=0.3$ GeV, the spectator quark mass $m_s=0.792$ GeV, and the $\pi^0$ mass $M=0.135$ GeV. The remaining spectator model parameters, $g_q$, $\lambda$, $a$, and $b$, are determined by fitting to the MAPFF set of $D_1$. We evolve the model result of $D_1(z)$ from the initial scale $\mu_0=1$ GeV to $Q=2.8$ GeV at NLO and fit each replica of the MAPFF set of $D_1^{u\rightarrow\pi^0}$, with the fit results shown in Fig.~\ref{fig:fitD1-MAPFF}, where the uncertainty bands are given at $68\%$ confidence level. The model parameter values are $g_q=4.473^{+4.528}_{-0.871}$, $\lambda=1.111^{+0.081}_{-0.199}$ GeV, $a=-1.119^{+0.668}_{-0.518}$, and $b=-0.697^{+0.049}_{-0.051}$.

\begin{figure*}[htp]
  \centering
 \includegraphics[width=0.95\linewidth]{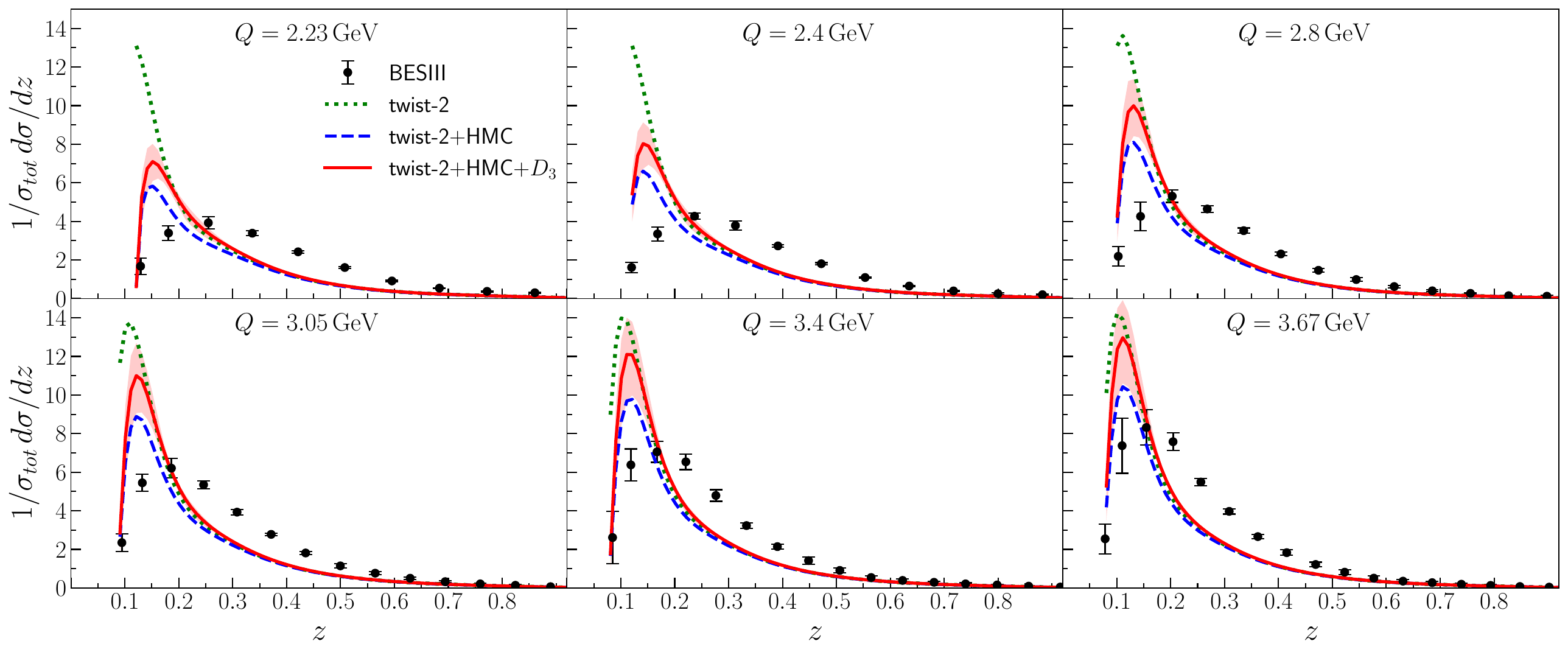}\\
  \caption{Normalized differential cross sections from the twist-2 contribution (green dotted curves), those including hadron mass correction (blue dashed curves), and those also including the $D_3$ contribution (red solid curves), are compared with the BESIII data for $e^+ e^- \rightarrow \pi^0 + X$~\cite{BESIII:2022zit}. The bands represent the uncertainty of the estimated $D_3$ contribution propagated from the model parameters.}\label{fig:fixQ-MAPFF}
\end{figure*}

\begin{figure*}[htp]
  \centering
 \includegraphics[width=0.95\linewidth]{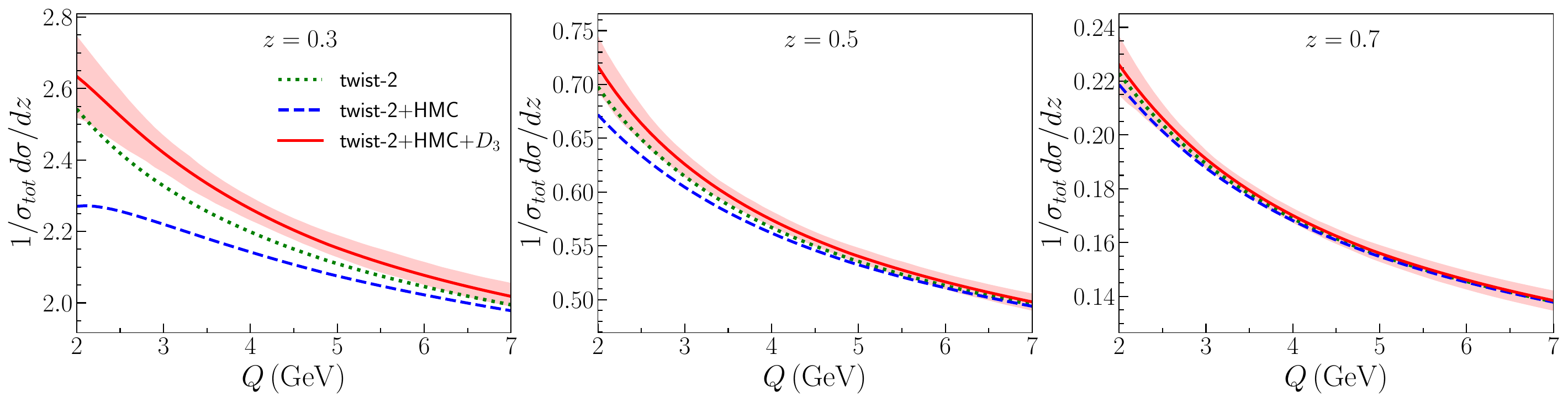}\\
  \caption{Dependence of the normalized differential cross section on the center-of-mass energy $Q$ at fixed $z=0.3$, $z=0.5$, and $z=0.7$, from the twist-2 contribution (green dotted curves), those including the hadron mass correction (blue dashed curves), and those also including the $D_3$ contribution (red solid curves). The bands represent the uncertainty propagated from the model parameters to the estimation of the $D_3$ contribution.}\label{fig:fixz-MAPFF}
\end{figure*}

After integrating Eq.~\eqref{f:crossparton4} over $\cos\theta$, the cross section for inclusive annihilation up to twist-4, normalized to the total cross section $\sigma_{tot}(e^+ e^- \rightarrow X)$, can be expressed as
\begin{align}
    \frac{1}{\sigma_{tot}} \frac{d\sigma_U}{dz}
    &=\frac{\beta}{\displaystyle{\sum_a} e_a^2} \sum_a e_a^2\bigg[D_1 + \beta^2 \frac{4\kappa_M^2}{z^2}D_3\nonumber\\
    &-\frac{4\kappa_M^2}{z}(D_{3dd}-G_{3dd})-\frac{\kappa_M^2}{z}D_{4q}\bigg],\label{f:normalized-cross-section}
\end{align}
where we use the subscript $U$ to indicate the unpolarized part of the cross section.
In this work, we evaluate only the $D_3$ term within the spectator model for simplicity, since it is generated by the quark-quark correlator in the same model setup as $D_1$. We note that this restriction is not an assumption that $D_3$ dominates over the other twist-4 FFs. The flavor sum runs over $u$, $d$, $\bar u$, and $\bar d$ flavors with the isospin and charge conjugation relations utilized. The strange contributions are omitted for simplicity in the present calculation. The modeled $D_3$ is evaluated at $Q=2.8$~GeV and is not evolved as the hard scale $Q$ is varied. As shown in Eq.~\eqref{f:normalized-cross-section}, the $Q$ dependence of the $D_3$ contribution mainly arises from the explicit $1/Q^2$ factor. The following numerical estimates adopt a fixed $D_3$ input without dynamical twist-4 evolution. The numerical results below are therefore regarded as an illustrative estimate of the potential contribution from twist-4 terms, rather than a complete twist-4 phenomenological calculation.

\begin{figure*}[htp]
  \centering
 \includegraphics[width=0.95\linewidth]{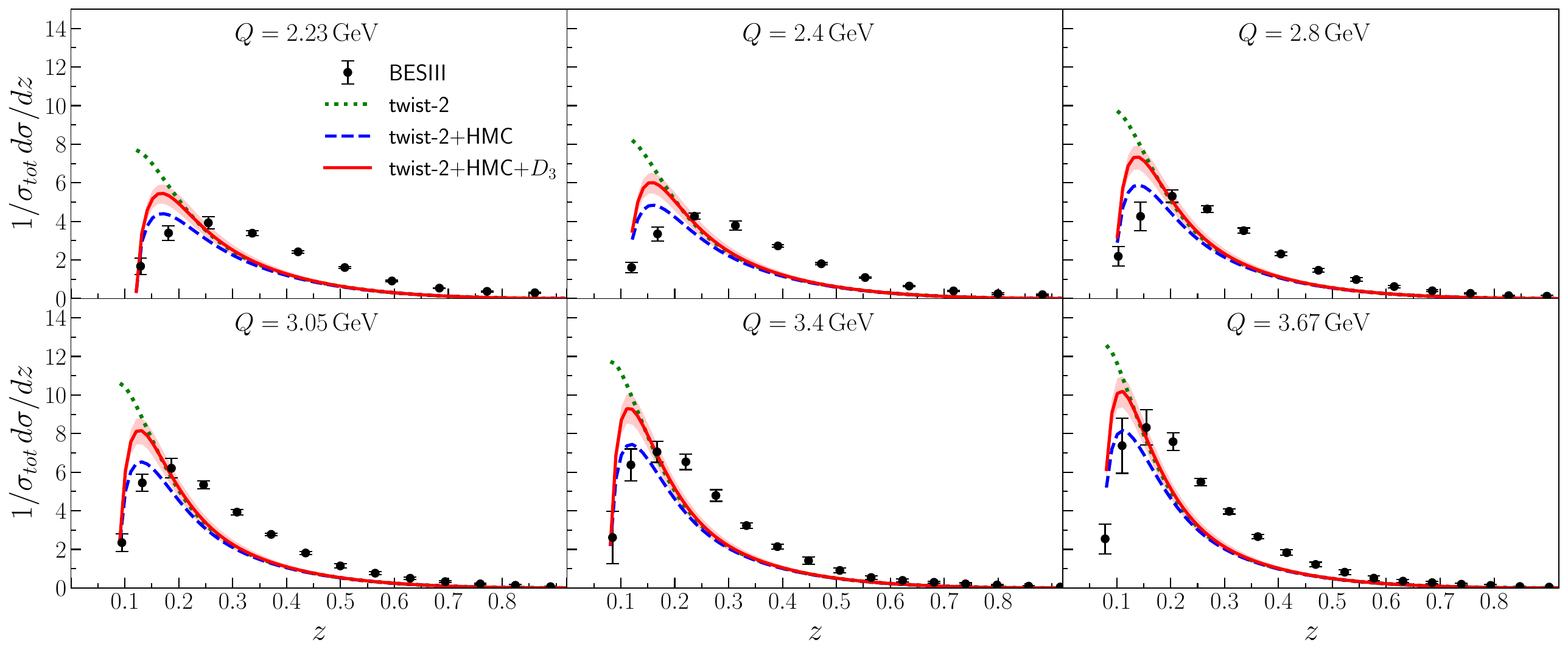}\\
  \caption{Normalized differential cross sections obtained using the leading-twist FF $D_1$ for $\pi^0$~\cite{Li:2024etc} as the leading-twist input and the modeled $D_3$ contribution evaluated with the spectator model parameters determined from the corresponding $D_1$ fit.}\label{fig:fixQ-npc}
\end{figure*}

For the normalized differential cross section containing only the twist-2 contribution, we directly adopt the MAPFF parametrization~\cite{AbdulKhalek:2022laj} and subsequently include the hadron mass correction and the estimated $D_3$ contribution. As illustrated in Fig.~\ref{fig:fixQ-MAPFF}, we present the corresponding numerical results. The green dotted curves represent the contribution from the leading-twist unpolarized FF only, labeled by twist-2; the blue dashed curves correspond to the contribution with the hadron mass correction, labeled by twist-2+HMC; and the red solid curves indicate the contribution including both the hadron mass correction and the estimated $D_3$ contribution, labeled by twist-2+HMC+$D_3$.
In contrast to the twist-2 results, which increase rapidly in the low $z$ region, the results are suppressed by the HMC at small $z$ and exhibit a peak.
When the $D_3$ contribution is included in addition to the twist-2 contributions and hadron mass corrections, the normalized differential cross section exhibits an enhancement relative to the result obtained without the $D_3$ term. The uncertainty band in Fig.~\ref{fig:fixQ-MAPFF} is obtained by propagating the spectator model parameter sets fitted to the MAPFF replicas to the $D_3$ contribution and normalized SIA cross section. With the constituent quark, spectator, and pion masses held fixed in the adopted spectator model, the uncertainty bands shown in Figs.~\ref{fig:fixQ-MAPFF}, \ref{fig:fixz-MAPFF}, and~\ref{fig:fixQ-npc} only include the propagated parameter uncertainty and exclude uncertainties associated with the model form, omitted twist-4 terms, and the truncation of perturbation series, thereby being understood as a partial assessment of the theoretical uncertainty in the twist-4 contribution. Around the relatively low-$z$ region, $0.1<z<0.2$, where the hadron mass correction leads to a kinematic suppression, the twist-4 FF $D_3$ contribution produces an enhancement of approximately $22\%\pm 14\%$. Within the uncertainty band, the modeled $D_3$ contribution can partially improve the agreement with the BESIII data in the $0.15<z<0.5$ region, while noticeable discrepancies remain over the full measured region. We note that as $z$ approaches the lower limit $z_{\min}=2M/Q$, the reliability of the twist expansion may be limited by higher powers of $M/zQ$ and possible threshold effects.
On the other hand, the modification induced by the modeled $D_3$ contribution decreases as $Q$ increases, which is qualitatively consistent with the expectation that twist-4 contributions are suppressed by an additional power of $1/Q^2$. Consequently, this estimate suggests that the twist-4 FF $D_3$ contribution is non-negligible in low-energy SIA processes, but a complete assessment requires the inclusion of all other twist-4 FFs.
We note that MAPFF parametrization~\cite{AbdulKhalek:2022laj} does not include higher-twist corrections, and low energy data are not used for global analysis. This may be the reason that theoretical numerical estimates cannot perfectly describe the BESIII experimental data~\cite{BESIII:2022zit,BESIII:2024hcs}. 

\begin{figure}[htp]
  \centering
 \includegraphics[width=0.8\linewidth]{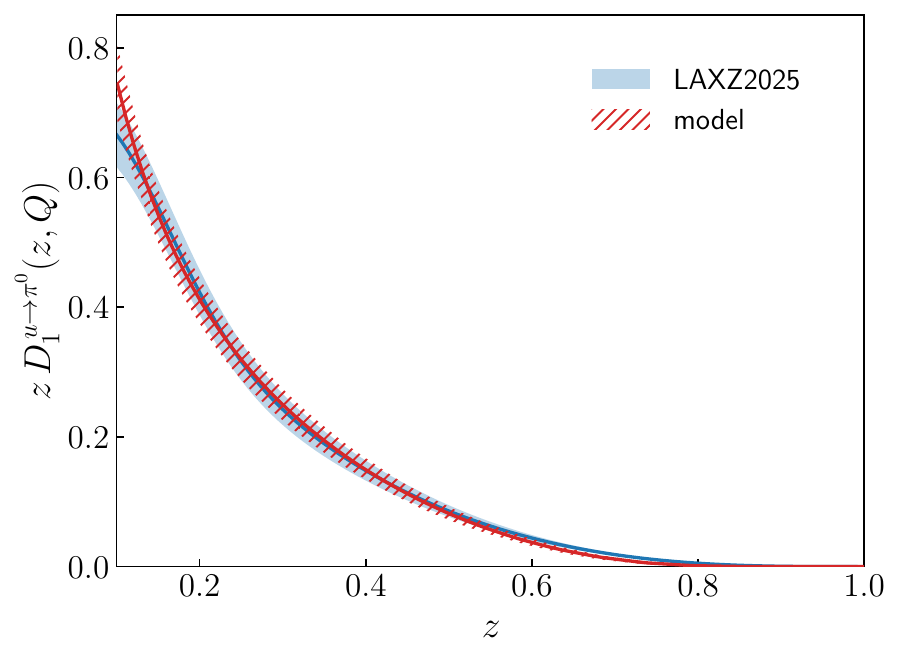}\\
  \caption{Comparison of the spectator model fit for $D_1$ with the leading-twist $\pi^0$ fragmentation function in Ref.~\cite{Li:2024etc} at $Q=2.8$ GeV.}\label{fig:fitD1-NPC}
\end{figure}

In phenomenological extractions of leading-twist FFs, higher-twist effects can be partially absorbed into the extracted leading-twist results if higher-twist terms are not treated as explicit contributions. To reduce the risk of counting such a contribution twice, we use the leading-twist $\pi^0$ FF $D_1$ in Ref.~\cite{Li:2024etc}, in which a higher-twist contribution was included in the fit to the BESIII data. Following the same fitting procedure used for the MAPFF replica analysis, we refit the spectator model $D_1$ to this alternative leading-twist FF set, as shown in Fig.~\ref{fig:fitD1-NPC}. The resulting model parameters are $g_q=5.011^{+0.146}_{-0.204}$, $\lambda=1.712^{+0.372}_{-0.229}$ GeV, $a=-0.441^{+0.218}_{-0.162}$, and $b=0.342^{+0.382}_{-0.189}$. The corresponding parameter ensemble is then used to evaluate the modeled $D_3$ contribution. As shown in Fig.~\ref{fig:fixQ-npc}, the estimated $D_3$ term still gives a visible modification in the intermediate-$z$ region. A complete separation of leading-twist and higher-twist FFs, however, requires a dedicated global analysis including both the leading-twist and twist-4 FFs.

Considering the future electron–positron collider at these energy scales, such as the Super Tau Charm Facility (STCF)~\cite{Ai:2025xop,Achasov:2023gey}, which is expected to operate with center-of-mass energies ranging from 2 to 7 GeV, we also illustrate the dependence of the normalized differential cross section on $Q$ within this range in Fig.~\ref{fig:fixz-MAPFF}, at fixed values of $z = 0.3$, $z = 0.5$, and $z = 0.7$.
When $z$ is fixed, the difference induced by the modeled $D_3$ contribution decreases with increasing $Q$, consistent with the physical expectation that twist-4 effects are suppressed at higher energies.
On the other hand, the blue and green curves tend to converge as $Q$ increases, indicating that the contribution from hadron mass corrections becomes negligible at relatively high energies, i.e., as $M/Q \to 0$. This behavior further suggests that more scrutiny is required on the higher-twist contributions.

\section{Summary}\label{sec:summary}

In this paper, we have established a comprehensive theoretical framework for the SIA process, $e^+ e^- \to h X$, incorporating higher-twist contributions up to twist-4 within the collinear factorization formalism. By performing a systematic collinear expansion of the hadronic tensor, we derived the complete set of structure functions expressed in terms of gauge-invariant FFs, considering both beam and hadron polarizations. Our derivation explicitly accounts for contributions arising from two-parton quark-quark, three-parton quark-gluon-quark, and four parton quark-gluon-gluon-quark and four-quark correlators. We have further verified that the derived hadronic tensor satisfies electromagnetic current conservation, $q_\mu W^{\mu\nu} = 0$, at the twist-4 level, ensuring the gauge invariance and theoretical consistency of the framework.

To assess the phenomenological impact of these power corrections, we performed a numerical estimation for $\pi^0$ production. We employed a spectator model to evaluate the twist-4 unpolarized fragmentation function $D_3$, which is currently unconstrained by global analyses. We specifically investigated the interplay between kinematic hadron mass corrections and the twist-4 FF $D_3$ contribution. Our numerical estimates indicate a possible interplay in the low-$z$ region: while HMC leads to a kinematic suppression of the differential cross section, the estimated $D_3$ contribution may induce an enhancement.

Comparing our numerical results with recent experimental data from the BESIII Collaboration at center-of-mass energies $Q \in [2.0, 3.67]$ GeV, we observed that the inclusion of the modeled $D_3$ contribution partially improves the description of the data in the region $0.15 < z < 0.5$. The analysis of the $Q$-dependence shows that this contribution follows the expected additional $(1/Q^2)$ power behavior, being suppressed at high energies, but is potentially non-negligible at the intermediate energy scales, such as at BESIII and the proposed STCF. These findings suggest that higher-twist effects may contribute to the tension between leading-twist predictions and low-energy SIA data. While only one twist-4 contribution is estimated in this study, future high-precision measurements at these energies would benefit from a systematic examination of higher-twist FFs in a consistent framework to improve the reliability of the extraction of FFs and the understanding of the hadronization process.

\acknowledgments{
We thank Ke Yang and Xiaoyan Zhao for valuable discussions.
This work was supported by the National Key R\&D Program of China No.~2024YFA1611004, the National Natural Science Foundation of China (Grants No.~12321005, and No.~12405103), and the Youth Innovation Technology Project of Higher School in Shandong Province (No.~2023KJ146).
}

\section*{Data availability}

The data of numerical results in Figs.~\ref{fig:fitD1-MAPFF}--\ref{fig:fixQ-npc} are available in~\cite{supplemental}.

\end{document}